\documentclass[epj]{svjour} 
\usepackage{graphicx}
\usepackage[]{epsfig}
\usepackage[dvips]{color}
\usepackage{multirow}

\newcommand{\remove}[1]{{}}

\usepackage{bm}
\DeclareBoldMathCommand{\bfmu}{\mu}

\begin{document}
\sloppy

\title{Magnetic interactions of cold atoms with anisotropic conductors}

\author{T David\inst{1} \and Y Japha\inst{1} \and V Dikovsky\inst{1} \and R Salem\inst{1} \and C Henkel\inst{2} \and R Folman\inst{1}}
\institute{Department of Physics, Ben-Gurion University, Be\'er Sheva 84105, Israel, \email{taldavid@bgu.ac.il} \and Institut f\"ur Physik und Astronomie, Universit\"at Potsdam, 14476 Potsdam, Germany} 
\date{\today}

\titlerunning{Magnetic interactions of cold atoms with anisotropic conductors} 
\authorrunning{T David \itshape et al. \upshape} 

\abstract{We analyze atom-surface magnetic interactions on atom chips where the magnetic trapping potentials are produced by current carrying wires made of electrically anisotropic materials.
We discuss a theory for time dependent fluctuations of the magnetic potential, arising from thermal noise
originating from the surface.  It is shown that using materials with a large electrical anisotropy results in  a considerable reduction of heating
and decoherence rates of ultra-cold atoms trapped near the surface, of up to several orders of magnitude.  The trap
loss rate due to spin flips is expected to be significantly reduced
upon cooling the surface to low temperatures.  In addition, the electrical anisotropy
significantly suppresses the amplitude of static spatial potential corrugations due to
 current scattering within imperfect wires. Also the shape of the corrugation pattern depends on the electrical anisotropy: the preferred angle of the scattered current wave fronts can be varied over a wide range. Materials, fabrication, and experimental issues are discussed, and specific candidate materials are suggested.
\PACS{{37.10.Gh}{Atom traps and guides} \and {39.25.+k}{Atom manipulation} \and {72.15.-v}{Electronic conduction in metals and alloys}  \and {03.75.-b}{Matter waves}}}

\maketitle 

\section{Introduction}
\label{Sect:INTRO}
The use of atom chips for trapping, cooling, manipulating and
measuring ultra-cold atoms near surfaces has attracted much attention
in recent years \cite{ronReview,Fortagh07}.  The monolithically integrated
micro-structures on the chip lead to tight potentials, which
can be tailored over length scales comparable to the atoms'
de-Broglie wavelength.
This has simplified the production of ultra-cold quantum degenerate gases
\cite{Reichel01b,TorontoFermi}, enabled high precision experiments
such as atom interferometry \cite{Reichel_clock,HDrfIFM},
and has established ultra-cold atoms as a probe of the nearby surface of
the atom chip \cite{HDMagMicroscope}.  To date, atom chip traps are
mostly magneto-static.  However the use of electrostatic, radiofrequency (rf), or light
potentials for atom manipulation on atom chips is also evolving
rapidly (e.g. \cite{Perrin,Prentiss,Lesanovsky07,Hasse,Wildermut,StarkChip}).
The chip platform is also considered a candidate for the
development of quantum technologies in the field of quantum
information processing and communication, as well as in interferometry
based sensors.  The advantages of being in close proximity to
the surface are hindered, however, by harmful processes originating from the
surface itself, which currently limit the capabilities of this
platform \cite{HenkelFundamentalLimits,HDFragmentation}.

Random motion of thermal electrons within the nearby surface (known as Johnson or
thermal noise) leads to magnetic field fluctuations, affecting the trapping potential and resulting in trap loss, heating, and decoherence.
As a general rule, the thermal noise increases as the atoms
approach the surface; therefore it sets limits to the confinement of
micro-traps and to the coherence time in sensitive interference
experiments \cite{ronReview,Fortagh07}.
The issue of trap loss due to spin flips has received the
largest attention so far. Several studies were done on characterizing
the lifetime of trapped atoms at different atom-surface distances, and
with different material surfaces
\cite{TubingenFragmentation,KetterleNoise,Cornell,TubingenNoise,Sussex_noise,Vuletic,ZhangNoise,ValeryAlloys},
and the theory was found to fit well to experimental data.
Recently, some theoretical work studied the reduction of
trap loss, when the surface is either made of certain metal alloys \cite{ValeryAlloys} or superconductors
\cite{Scheel,Rekdal,Scheel2,Rekdal2,Rekdal3,Rekdal4}, while cooling
the surface to cryogenic temperatures.  Very few experiments were done
on surface noise induced decoherence \cite{Reichel_clock,DecoherenceNote}.

Current scattering (change of current flow direction) from surface
or edge roughness or from bulk inhomogeneities leads to static spatial corrugations of the magnetic trapping potential \cite{HDFragmentation,Schumm}.
The consequent atomic density variation results in the fragmentation of the atom cloud into several separated components as it is brought closer to the surface.
Much work has been done theoretically and experimentally in order to
understand the origin of fragmentation
\cite{TubingenFragmentation,Schumm,Esteve,Lukin,aigner}. Improved fabrication has led to a considerable reduction of the corrugations, and schemes have been suggested for further improvement \cite{HDFragmentation,Trebbia}. Recently, highly organized corrugation patterns were observed and analyzed \cite{aigner,japha}, when a 1D cloud was scanned across lithography patterned wires of different
characteristics.
A prominent feature, observed for different types of
wires, were long range wave fronts along which current scattering is maximal, oriented at $\pm 45^{\circ}$ to the wire axis. This was explained in terms of maximally scattering Fourier components of random conductivity fluctuations in the wire \cite{aigner,japha}.

In this paper we study both thermal noise and
static potential corrugation in the context of atom
chip conductors that are electrically anisotropic.
These  have different conductance along the different crystal axes,
contrary to the isotropic conductivity of normal metals.
The properties of such materials, including ruthenates, cuprates, graphites and others, have been studied thoroughly (see Table 1 below). Regarding thermal noise, we show here that electrically anisotropic materials
lead to a significant reduction of the heating and
decoherence rates.  Trap lifetimes can be increased by cooling the
chip surface to low temperatures.  With respect to potential
corrugation, we show that the orientation of current scattering wave fronts within wires can be
tuned by the electrical anisotropy. For
high anisotropy, fragmentation is expected to be strongly suppressed.

In section \ref{Sect:noise} we discuss the implications of using\linebreak
anisotropic materials on the thermal noise produced by a surface, and
analyze the resulting trap lifetime, heating, and decoherence processes.  We
then extend in section \ref{Sect:fragmentation} the formulation used
previously \cite{aigner,japha} to analyze potential corrugation
and fragmentation to the case of electrically
anisotropic materials.  Finally, in section \ref{Sect:mat_fab} we discuss
issues of materials and fabrication, important for the design of
experiments to test the theory, as well as for future engineering of quantum technology devices.

\section{Loss, heating, and decoherence due to thermal noise}
\label{Sect:noise}
The theory describing the affect of thermal noise on ultra cold atoms is now well established for isotropic materials (e.g. \cite{ronReview,HenkelFundamentalLimits,HenkelPotting,HenkelPottingWilkens,ScheelSpatialDecoherence}). Other models, based for example on non-local electrodynamics, have also been developed \cite{Horovitz}. In this section we derive and discuss a generalized theory of thermal noise coupled to magnetically trapped atoms,
which includes the case of electrically anisotropic materials. Our method incoherently adds up elementary current sources in the material, within the quasi-static approximation~\cite{HenkelPotting}.
Although approximate (up to corrections on the order of a factor 2), this approach can be used to obtain closed-form solutions
for many wire geometries. This is in contrast to methods based on Green functions, which are
accurate but limited to simple geometries such as a half-space or
a laterally infinite layer. It should be noted that the quasi-static approximation is valid
when the skin depth $\delta=\sqrt{2/ \sigma_{0} \mu_0 \omega}$ (where $\sigma_0$, $\mu_0$, and $\omega$ are the electrical conductivity, permeability of free space, and the radiation frequency, respectively) is much
larger than both the atom-surface distance $d$ and the wire thickness
$H$.  At room-temperature the skin depth of Au at $1\,{\rm MHz}$
is on the order of $70 \mu m$. Electrically anisotropic materials usually have a much
higher resistivity, and their skin depths are orders of magnitude larger at the same frequencies
(e.g. for graphite $\delta\approx 1mm$, see Table 1 below).
The quasi-static approximation thus applies even better
to the latter. This approximation has been corroborated
to a high degree of accuracy in experiment~\cite{ValeryAlloys,Vuletic}
and theory~\cite{Zhang07}.

\subsection{Physical processes}
\label{Sect:rates}
We now introduce the different physical processes resulting from the
coupling of thermal noise to the trapped atoms.  We show that the important quantity is the noise power spectrum, and calculate
the latter in the following section.  The thermal radiation couples 
to the atoms'
magnetic moment by the Zeeman interaction $V\left(\vec{x},t\right)=-
\bfmu \cdot \vec{B}\left(\vec{x},t\right)$, where $\bfmu$ is 
the magnetic dipole moment operator.  This operator can
be written for convenience as $\bfmu = \mu_B g_F {\bf F}$, with $\mu_{\rm
B}$ Bohr's magneton, $g_F$ the Land\'e factor of the appropriate
hyperfine level, and ${\bf F}$ the total spin operator.  As the magnetic field
noise is random in time and space it averages to zero, and its effect
is expressed through its two-point correlation function $\left\langle
V\left(\vec{x}_1,t\right) V\left(\vec{x}_2,t'\right) \right\rangle$,
between different points in time $t,t'$ and space
$\vec{x}_1,\vec{x}_2$.

The field and magnetic moment being vectors, one actually needs the 
cross-correlations between their components to characterize
the different processes induced by the noise (spin flips, heating, 
and decoherence). It follows from time-dependent perturbation theory 
that the rate for a transition from an initial state
$\left|0\right\rangle$ to a final state $\left|f\right\rangle$ of the
system is given by \cite{LandauLifshitz}
\begin{eqnarray}
\Gamma_{0 \to f} & = & \frac{1}{\hbar^2}\sum_{i,j}\int\!{\rm d}(t - t')\,{\rm e}^{{\rm i}\omega_{0f}(t - t')}\times
\label{Eq:Fermi} \\
& \times & \left\langle \langle 0 | \mu_i B_i( {\bf x}, t ) | f \rangle \langle f | \mu_j B_j( {\bf x}, t' ) | 0 \rangle \right\rangle \nonumber
\end{eqnarray}
where the indices $i, j$ label the Cartesian components of the 
magnetic field. The argument ${\bf x}$ of the magnetic field in Eq. 
(\ref{Eq:Fermi}) is the atomic position operator. The matrix elements 
thus involve spatial (overlap) integrals over the spatial wave 
function and magnetic field (e.g.\ see Eq. (\ref{Eq:heatingRate}) 
below). These integrals can be written in terms of the
two-point correlation function of the magnetic field. The integral of 
this correlation function over the time difference $(t - t')$ is 
related to the cross-spectral density $S_B^{ij}( {\bf x}_1, {\bf 
x}_2; \omega_{0f} )$ of the magnetic fluctuations at positions ${\bf 
x}_1, {\bf x}_2$, and at the transition frequency $\omega_{0f}$ \cite{HenkelFundamentalLimits,Mandel}. This 
is defined as
\begin{eqnarray}
\lefteqn{S^{ij}_B\left(\vec{x}_1,\vec{x}_2;\omega\right)\equiv}  
\label{Eq:PowerSpectrumDefinition} \\
&& \equiv \int_{-\infty}^{\infty} {\rm d} \left(t-t'\right) e^{-{\rm 
i} \omega(t-t')}\left\langle 
B_i\left(\vec{x}_1,t\right)B_j\left(\vec{x}_2,t'\right)\right\rangle 
\nonumber, 
\end{eqnarray}
or equivalently by
\begin{equation}
\left\langle 
B_i^*\left(\vec{x}_1,\omega\right)B_j\left(\vec{x}_2,\omega 
'\right)\right\rangle = 2 \pi \delta \left(\omega - \omega 
'\right)S_B^{ij}\left(\vec{x}_1,\vec{x}_2;\omega\right).
\label{Eq:spectralDensity}
\end{equation}

In cases where the spatial degrees of freedom can be considered 
classically, ${\bf x}$ in Eq. (\ref{Eq:Fermi}) can be taken as the coordinate of the 
atoms, giving rise to a position-dependent transition rate between
spin states. Instead 
of the wave function overlap integrals, the coordinates are then 
taken as $\vec{x}_1=\vec{x}_2$. For calculating at the trap center 
one takes $\vec{x}_1=\vec{x}_2=\vec{r}$ \cite{Note:TrapInhomogeneity}.

The noise power spectrum $S_B^{ij}( {\bf x}_1, {\bf x}_2; \omega_{0f} 
)$ is quite flat at all relevant frequencies which correspond for 
example to transitions between Zeeman magnetic sub-levels (rf), or 
the trap vibrational states (Hz to kHz range). Hence a low-frequency 
limit $\omega_{0f} \rightarrow 0$ can be taken. If this spectrum is 
not flat in the relevant frequency range, the time dependence of loss 
and decoherence processes is more complicated and does not reduce to 
a simple rate \cite{HenkelFundamentalLimits}.

Spin flips are transitions from trapped ($|0\rangle$) to untrapped 
($|f\rangle$) internal states. These states are eigenstates of the 
spin operator component parallel to the quantization axis, 
defined by the static trapping field at the bottom of the trap. 
Due to the form of the dipole moment operator $\mu_i$, 
the Zeeman interaction can induce a spin flip 
only when the direction of the magnetic field fluctuation is 
perpendicular to the quantization axis.
Hence, in this case, we can rewrite (\ref{Eq:Fermi}) as
\begin{equation}
\gamma_{{\rm spin~ flip}} = \sum_{l,m=\perp} \frac{\mu_l \mu_m 
}{\hbar^2}S_{B}^{lm}\left(\vec{r};\omega_{0f}\right),
\label{Eq:spinfliprate}
\end{equation}
where we sum over the perpendicular components $l,m$ only, and take 
the matrix elements $\mu_l = \mu_B g_F \langle 0 | F_l | f \rangle$ 
of the dipole moment between the states $|0\rangle$ and $|f\rangle$.
The spin flip rate can be measured from the lifetime of a magnetic
micro-trap, and by varying .the trap distance, one can discriminate loss due to surface induced magnetic field fluctuations, against trap loss of different origin~\cite{Cornell,TubingenNoise,Sussex_noise,Vuletic,ZhangNoise}.

Decoherence of a quantum superposition state can occur without 
changing the occupation of the states involved in the superposition, 
affecting only the phase. We distinguish between spin decoherence and 
spatial decoherence. The former involves the change of relative phase 
of two internal states ($|1\rangle$ and $|2\rangle$) in a 
superposition at the same spatial location ${\bf r}$, while the latter involves 
the change of relative phase of two spatially separated components of 
the atom cloud (positions ${\bf x}_{1}$ and ${\bf x}_{2}$),
trapped in the same internal state. 
Here we consider only `classical decoherence' in which the phase 
change arises from the fluctuations in the Zeeman shift. Such shifts occur, to lowest order, 
for magnetic field fluctuations parallel to the atom's magnetic 
dipole moment. Hence for both types of decoherence processes we need 
only the parallel component of the same power spectrum appearing in 
the spin flip case (Eq. (\ref{Eq:spinfliprate})). The spin 
decoherence rate can thus be written as
\begin{equation}
\gamma_{{\rm spin~ decoherence}} = \frac{\Delta 
\mu_\|^2}{2\hbar^2}S_\|\left(\vec{r};0\right), 
\label{Eq:spinDecoherenceRate}
\end{equation}
with $\Delta \mu_\| = \left\langle 2 \left| \mu_\| \right| 
2\right\rangle - \left\langle 1 \left| \mu_\| \right| 1 
\right\rangle$ being the differential magnetic
moment of the two trapped states, and $S_\|\left(\vec{r};0\right)$ 
the low-frequency limit parallel component of the
noise spectrum, at the trap center $\vec{r}$. This decoherence rate 
can be measured from the reduced contrast
of interference fringes in a series of Ramsey spectroscopy 
experiments, performed as a function of the atom-surface distance.

For the case of a spatially separated superposition state, the rate of decoherence of the relative phase of the atomic states between two points $\vec{x}_1,\vec{x}_2$ involves the correlation function of the difference in the magnetic fields $B_\|( \vec{x} _1, t ) -B_\|(\vec{ x}_2, t' )$, and can be written as
\begin{equation}
\gamma_{{\rm spatial~decoherence}} = \frac{\mu_\|^2}{2\hbar^2} \left[S_{11}+S_{22}-2S_{12}\right],
\label{Eq:spatialDecoherenceRate}
\end{equation}
where we denote for convenience $S_{ij}=S_\|(\vec{x}_i,\vec{x}_j;\omega \rightarrow 0)$, and assume a correlation spectrum symmetric in $\vec{x}_1$, $\vec{x}_2$ and flat in frequency. We again take only the field components parallel to the quantization axis, that shift the relative phase between the two parts of the wave function. $\mu_\|$ is the magnetic moment matrix element of the single internal state. The low-frequency limit is valid here as the ``scattering cross section" for transitions involved in the decoherence process is in practice independent of the frequency \cite{HenkelFundamentalLimits,ScheelSpatialDecoherence}.

This decoherence rate can be measured by studying the interference pattern 
contrast reduction in double-well experiments, where the atom cloud 
is separated in two parts, which are then held at a fixed separation 
for a given time. The interference pattern is obtained by overlapping 
the cloud parts upon release from the trap. The interference contrast 
is directly related to the product of split time and decoherence 
rate; it can be measured as a function of the atom-surface distance 
by repeating the measurement at different heights.

Finally, heating of the trapped atoms (as a result of exciting
external degrees of freedom while retaining the same internal state) 
can also be caused by the coupling to the thermal radiation. 
The transition rate of atoms initially 
in the ground vibrational state $\left|0\right\rangle$
to higher states $\left|f\right\rangle$ with energy splitting
$\hbar \omega_{0f}$ is of the form 
\cite{HenkelFundamentalLimits}
\begin{equation}
\Gamma_{0 \rightarrow f} = \frac{\mu_\|^2}{\hbar^2} \int {\rm 
d}\vec{x}_1 {\rm d}\vec{x}_2 M_{f0}^*(\vec{x}_1) M_{f0}(\vec{x}_2) 
S_\|\left(\vec{x}_1,\vec{x}_2; \omega_{0f} \right), 
\label{Eq:heatingRate}
\end{equation}
where we find once more the spin operator matrix element $\mu_\|$ in 
the direction parallel to the quantization axis, 
and now also the wave functions of 
the levels involved in the transition $M_{f0}( {\bf x} ) = 
\psi_f^*(\vec{x})\psi_0(\vec{x})$.
The spatial integration 
here provides a probe of the spatial correlation of the magnetic 
field noise. In practice, it is 
enough to consider transitions from the ground state to either of the 
first two excited levels \cite{ronReview}. 

Thus, we see that the important term common to all rates is the 
spectral density of the magnetic field fluctuations or power spectrum 
$S_B^{ij}\left(\vec{x}_1,\vec{x}_2;\omega\right)$. This quantity 
holds all of the relevant information about the magnetic field 
fluctuations leading to the harmful processes, while the other terms 
in each of the rates describe the coupling of the noise to the atoms 
through the magnetic dipole moment. Furthermore, we see that a 
measurement of either the spin or spatial decoherence rates, or of 
the heating rate, will give strong indications to any of the other 
two of these three processes, as they all depend on the noise power 
of the same field component. 

\subsection{Calculation of the noise power spectrum}
\label{Sect:powerSpectrum}

As presented in the previous section, the noise power spectrum is related to the cross correlation function of the magnetic field fluctuations (Eqs. (\ref{Eq:PowerSpectrumDefinition}),(\ref{Eq:spectralDensity})). Before calculating this correlation function, we define the coordinate system
such that the wire length $L$ is along the $\hat{x}$ direction, its
width $W$ along the $\hat{y}$ direction, and its thickness $H$ along
$\hat{z}$.  As in typical magnetic traps a bias field
polarizes the atoms along the wire, the quantization axis in our analysis is along the $\hat{x}$ axis.

We can make a distinction between two types of\linebreak anisotropic
materials. Materials with a `layered conductance' are relatively good conductors
along two axes and have one axis of bad conductance: $\sigma_{yy} \ll
\sigma_{xx} \sim \sigma_{zz}$. We always assume that the
wire is aligned to one axis of good conductance, so that current may
flow easily and create a magnetic trap. Materials that have only one
direction of good conductance, $\sigma_{yy} \sim \sigma_{zz} \ll
\sigma_{xx}$ or $\sigma_{zz} \ll \sigma_{yy} \ll
\sigma_{xx}$, will be denoted as `quasi-1D conductors'. Specific materials are discussed in Sect. \ref{Sect:mat} and Table \ref{Tab:tablecaptionis}.

Turning now to the magnetic field fluctuations, we use the expression for the current cross correlation
function \cite{Rytov,LifshitzRef},
\begin{eqnarray}
\lefteqn {\left\langle j_i^*\left(\vec{x}_1,\omega\right)j_j\left(\vec{x}_2,\omega '\right)\right\rangle =} \nonumber \\
      & & 4\pi \hbar \epsilon_0 \omega^2
\overline{n}\left(\omega\right)
\delta\left(\omega - \omega '\right)
\,{\rm Im}\,\epsilon_{ij}\left(\vec{x}_1; \omega \right)
\delta\left(\vec{x}_1 - \vec{x}_2\right)  \label{Eq:Lifshitz},
\end{eqnarray}
where $\overline{n}\left(\omega\right) = 1 / ({e^{\frac{\hbar\omega}{k_{\rm B} T}}-1})$
is the Bose-Einstein occupation number, and where the dielectric tensor
\begin{equation}
\epsilon_{ij}(\omega)=\frac{i \sigma_{ij}}{\epsilon_0 \omega}
\label{Eq:epsilonSigma}
\end{equation}
 is proportional to the conductivity tensor for homogeneous media.
When the anisotropic crystal axes
are aligned with the wire, the latter is diagonal,
\begin{equation}
\hat{\sigma} = 
\bordermatrix{&  &  &  \cr
	 & \sigma_{xx} & 0 & 0 \cr
	 & 0 & \sigma_{yy} & 0 \cr
	 & 0 & 0 & \sigma_{zz} \cr}.\label{Eq:rhoDef}
\end{equation}

This property will be used below. We assume here that the current in the material responds locally to the electric field (Ohm's law), $j_i( {\bf x} ) = \sigma_{ij}( {\bf x} ) E_{j}( {\bf x} )$.
The vector potential and its correlation function in the quasi-static approximation are given by
\begin{eqnarray}
&\vec{A}\left(\vec{x},\omega\right) = \frac{\mu_0}{4\pi} \int {\rm d}\vec{x'}\frac{\vec{j}\left(\vec{x'},\omega\right)}{\left|\vec{x} - \vec{x'}\right|} \label{Eq:vector_potential} \\
&\left\langle A_i^*\left(\vec{x}_1,\omega\right)
A_j\left(\vec{x}_2,\omega '\right)\right\rangle
\propto \int_V {\rm d} \vec{x}'
\frac{\sigma_{ij}( \vec{x}' )
     }{ \left|\vec{x}_1 - \vec{x}'\right|\left|\vec{x}_2 - \vec{x}'\right|
     }
\label{Eq:vector_potential_correlation}.
\end{eqnarray}
Here $\vec{x}_1,\vec{x}_2$ denote the two spatial locations for which we take the correlation function, whereas $\vec{x}'$ is the integration variable, such that we sum the contribution from all points within the material volume $V$, which are at distances $\left|\vec{x}_i - \vec{x}'\right|$ ($i=1,2$) from the locations $\vec{x}_1,\vec{x}_2$. Note that this formula neglects jumps in $\vec{A}$ due to surface currents at the metal-vacuum interface. In order to calculate the correlation function of the magnetic field fluctuations, we take the curl of the vector potential correlation function, once with respect to $\vec{x}_1$ and once with respect to $\vec{x}_2$. Writing this in tensor form we get,
\begin{eqnarray}
\lefteqn{\left\langle B_i^*\left(\vec{x}_1,\omega\right) B_j\left(\vec{x}_2,\omega'\right)\right\rangle \propto}  \nonumber \\
& &\frac{1}{2}\int {\rm d}\vec{x'} \epsilon_{ilm} \epsilon_{jnp} \partial_{1,l} \partial_{2,n} \frac{\sigma_{mp}}{\left|\vec{x}_1 - \vec{x'}\right|\left|\vec{x}_2 - \vec{x'}\right|} \equiv B_{ij},\, \label{Eq:Bij1st}
\end{eqnarray}
using the Levi-Civita symbol $\epsilon_{ijk}$ and summing over repeated indices. We defined the integral holding the conductivity tensor and the geometry terms as $B_{ij}$ for convenience, and the symbol $\partial_{\alpha,l}$ ($\alpha = 1,2$) means a derivative with respect to $\vec{x_\alpha}$ in the direction of its $l$th-component. Performing the derivatives
we obtain
\begin{equation}
B_{ij} = \epsilon_{ilm}\epsilon_{jnp}\sigma_{mp}X_{ln},
\label{Eq:Bij_integral}
\end{equation}
where the geometry of the system enters through the quantity $X_{ln}$ (correcting a missing factor $1/2$ in Eq. (A.6) of
Ref. \cite{HenkelPotting}):
\begin{equation}
X_{ij} = \frac{1}{2}\int_V {\rm d} \vec{x'} \frac{\left(\vec{x}_1 - \vec{x'}\right)_i\left(\vec{x}_2 - \vec{x'}\right)_j}{\left|\vec{x}_1 - \vec{x'}\right|^3\left|\vec{x}_2 - \vec{x'}\right|^3}.
\label{Eq:Xij}
\end{equation}
We assume here a homogeneous medium (i.e., the components $\sigma_{ij}$ are
spatially constant within the material volume $V$).

Again limiting the discussion to
the 'aligned' case (\ref{Eq:rhoDef}), Eq. (\ref{Eq:Bij_integral}) is simplified, and in fact for every pair
of $i,j$ we need to sum only two integrals.  Considering the wire
geometry to be such that the atoms are located above the center of a
very long wire ($L \gg H,d$), the only non-zero elements are $B_{ii}$  ($i=x,y,z$) due
to symmetry.  Hence we obtain
\begin{eqnarray}
&B_{xx} = \sigma_{zz} X_{yy} + \sigma_{yy} X_{zz} \nonumber \\
& ~B_{yy} = \sigma_{zz} X_{xx} + \sigma_{xx} X_{zz}, \label{Eq:B_ii} \\
&B_{zz} = \sigma_{yy} X_{xx} + \sigma_{xx} X_{yy} \nonumber
\end{eqnarray}
and it is convenient to define
\begin{equation}
\tilde{Y}_{ij} \equiv B_{ij} / \sigma_{xx},
\label{Eq:Yij}
\end{equation}
assuming the good conductivity to be along $\hat{x}$.

We see that in contrast to the isotropic case, where one
has a single conductivity $\sigma_0$ for all field components $B_{ij}$,
here the conductivities $\sigma_{ii}$ give different weights to the
geometry-dependent factors $X_{ii}$.

We can now write the full expression for the power spectrum, using (\ref{Eq:spectralDensity}) and collecting all of the prefactors neglected in (\ref{Eq:vector_potential_correlation}),(\ref{Eq:Bij1st}) as
\begin{equation}
S_B^{ij}\left(\vec{x}_1,\vec{x}_2;\omega\right) = S_B^{bb}(\omega)\frac{3}{4\pi \omega/c}{\rm Im}\epsilon_{xx}\tilde{Y}_{ij},
\end{equation}
where, following \cite{HenkelPotting}, we have normalized the power spectrum to Planck's blackbody formula
\begin{equation}
S_B^{bb}(\omega)=\frac{\hbar \omega^3\bar{n}(\omega)}{3\pi \epsilon_0 c^5}.
\label{Eq:planck}
\end{equation}

As the relevant frequencies are low, the high temperature limit of the Planck function is applicable, and thus the expression for the power spectrum approaches
\begin{equation}
S_B^{ij}\left(\vec{x}_1,\vec{x}_2;\omega\right)= \frac{k_B T}{4\pi^2 \epsilon_0^2 c^4} \sigma_{xx} \tilde{Y}_{ij}.
\label{Eq:powerSpectrumFull}
\end{equation}

This expression has the same form as the result for the isotropic case \cite{HenkelPotting}, however here the tensor $\tilde{Y}_{ij}$ holds the anisotropic terms $B_{ij}$ in it.

\subsection{Trap loss due to spin flips}
\label{Sect:TrapLoss}

Examining Eq. (\ref{Eq:B_ii}), we see that $B_{xx}$ can be considerably reduced if it
involves two badly conducting axes, $\sigma_{yy}, \sigma_{zz}$ being
much smaller than $\sigma_{xx}$. This does not lead,
however, to a reduction of the spin flip rate, as the $B_{xx}$ noise
is parallel to the quantization axis.

For the two perpendicular components, $B_{yy}$ and $B_{zz}$,
we find that both mix the highly conducting $\sigma_{xx}$ and
the low conductivity terms.
 The geometrical factors $X_{ij}$ have been analyzed in
detail (see appendix A in \cite{ValeryAlloys}) for the case of
rectangular wires.  From this analysis it emerges that for
any reasonable wire geometry, all of the non-zero $X_{ij}$ factors are on the same order. Thus the main difference in the noise components is due to the difference in conductivity terms, where conductivity $\sigma_{xx}$ is dominant.  Consequently, the improvement to the trap lifetime using anisotropic materials at room temperature is expected to be at most on the order of $\sim\! 2$.

One could, of course, rotate the anisotropic crystal by $90^{\circ}$ such that the better
conducting axis is perpendicular to the wire. This
may significantly reduce the spin flip rate for some materials, as now the conductivity
along the quantization axis (along the wire) is smaller.
However, this is not practical if the wire is to be used
as a current carrying structure creating the magnetic trapping
potential, since the power consumption will be significantly increased.
This and other considerations on materials and experimental design will
be discussed in section \ref{Sect:mat_fab}.

\begin{figure}[h]
\centering
\includegraphics*[width=\columnwidth]{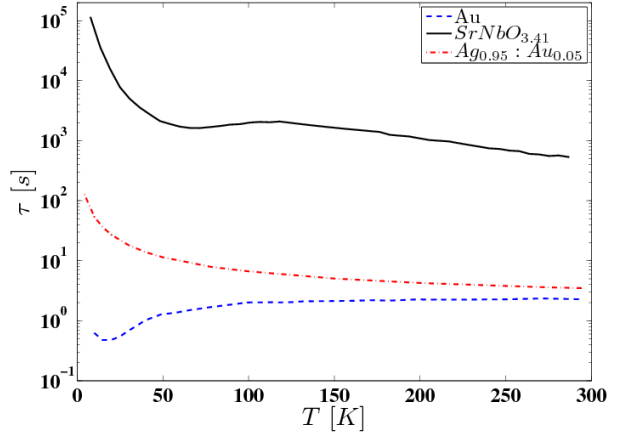}
\caption{Improved trap lifetime upon cooling of the surface.
Comparison of a standard Au wire (dahsed blue) with wires of similar geometry
made of an Ag:Au alloy (dashed-dotted red) \cite{ValeryAlloys} and the quasi-1D ${\rm SrNbO}_{3.41}$ (solid black). The long
lifetime for the anisotropic wire at room temperature is due to
its rather high residual (low-temperature) resistivity even along its best axis (oriented along the
wire). See section \ref{Sect:fab_noise} for the
implications of highly resistive wires in realistic experiments.
No limiting lifetime level due to collisions with background gas atoms is included. Losses due to this process are expected to be less significant in cryogenic experiments as typically the background pressure is such experiments is smaller. Wire dimensions are: width $W=10 \mu m$, thickness $H=2.15
\mu m$, and atom-surface distance $d = 5 \mu m$.
\label{Fig:coolingLifetime}}
\end{figure}

The trap lifetime may be improved nevertheless by cooling the anisotropic material
to cryogenic temperatures, as has been shown also for certain metal alloys
\cite{ValeryAlloys}. The resulting improvement in lifetime is shown in
Fig. \ref{Fig:coolingLifetime}, where we compare as an example the lifetime due to thermal
noise from an ${\rm SrNbO}_{3.41}$ wire (a quasi-1D material, see Table 1 \cite{TableSrNbO}) to that from
a Au and an Ag:Au alloy wires. We see that for ${\rm SrNbO}_{3.41}$ an improvement of two
orders of magnitude in lifetime is expected upon cooling. This is due to
the behavior of the product $T  \cdot \sigma_{xx}(T) $ (Eq. (\ref{Eq:powerSpectrumFull}))
for this and similar materials. We note that although at low temperatures the skin
depth is decreased due to the smaller resistivity of the material, the
quasi-static approximation is still valid. For a detailed discussion see \cite{ValeryAlloys}.

\subsection{Heating and decoherence}
\label{Sect:HeatindAndDecoherence}

\begin{figure}
\centering
\includegraphics*[width=\columnwidth]{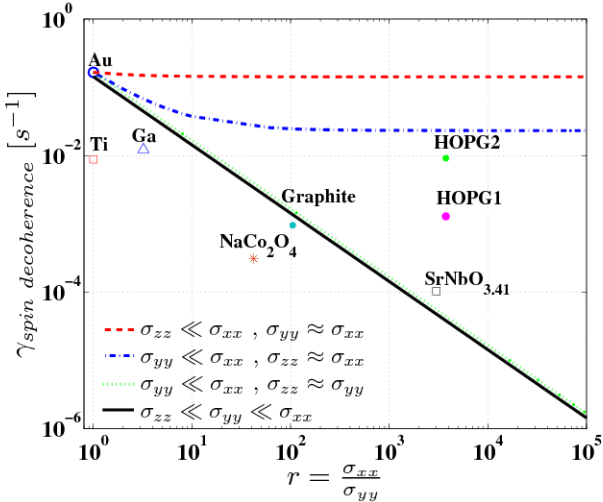}
\caption{Lines: Spin decoherence rate $\gamma_{{\rm spin~ decoherence}}$ as a function of electrical anisotropy $r=\sigma_{xx}/\sigma_{yy}$, for layered and quasi-1D conducting materials. Wire parameters were $d=5\mu m$, $W=10\mu m$, $H=2.15\mu m$, and surface temperature of $T=300 K$. For these lines, the good conductivity along the wire was assumed to be identical to that of Au. For layered materials having the badly conducting axis along the wire thickness (dashed red), only a slight improvement is gained, and the dependence on the anisotropy is negligible. If the badly conducting axis is along the width of the wire (dashed-dotted blue), the improvement is more pronounced, but still saturates at relatively low anisotropy. For quasi-1D materials (dotted green - quasi-1D with both low conductivity terms of the same order; solid black - the more extreme case having $\sigma_{zz} \ll \sigma_{yy} \ll \sigma_{xx}$, where we assumed $\sigma_{zz} \approx \sigma_{yy} / r$), the suppression is much more significant. For high anisotropy the scaling is linear with the anisotropy. Points: Examples of specific materials, not normalized to Au (see Table \ref{Tab:tablecaptionis}). \label{Fig:spinDecoherence}}
\end{figure}

In contrast to the case of spin flips, the
decoherence and heating rates,
Eqs. (\ref{Eq:spinDecoherenceRate},\ref{Eq:spatialDecoherenceRate},\ref{Eq:heatingRate}),
depend only on the noise in the parallel field component  $B_{xx}$.  As shown in
section \ref{Sect:TrapLoss}, this component may be strongly reduced for
highly anisotropic materials. To quantify this, consider
the ratio
\begin{equation}
\frac{B_{xx}^{{\rm aniso}}}{B_{xx}^{{\rm iso}}} = \frac{\sigma_{zz}X_{yy} + \sigma_{yy}X_{zz}}{\sigma_{xx}\left(X_{yy} + X_{zz}\right)},  \label{Eq:noiseSuppressionRatio}
\end{equation}
where the reference level is an isotropic conductivity
set to the
`good axis', $\sigma_{0} = \sigma_{xx}$.
The possible improvements depend on the relative
magnitudes of the anisotropic conductivities. Four cases can be
distinguished, as illustrated in Fig. \ref{Fig:spinDecoherence}.
For materials with layered conductance, the worst choice is to have
the second good conducting axis in the chip plane, along the wire's
width: $\sigma_{yy} \sim \sigma_{xx} \gg \sigma_{zz}$ (dashed red line).
The ratio~(\ref{Eq:noiseSuppressionRatio}) then tends to
$\left(1 + X_{yy} / X_{zz} \right)^{-1}$ which is not significantly
smaller than unity and where the conductivity anisotropy $\sigma_{xx} / \sigma_{yy}$ actually does not enter.
With the other choice, having the badly conducting axis along the wire
width (dashed-dotted blue), we get a reduction of about one order of magnitude for a small
wire geometry. Materials that are quasi-1D conductive have a much larger potential for noise
suppression: for comparable `bad axes', $\sigma_{yy} \sim \sigma_{zz}
\ll \sigma_{xx}$, Eq. (\ref{Eq:noiseSuppressionRatio}) scales inversely with the
anisotropy ratio $r=\sigma_{xx} / \sigma_{yy}$ which may be very large. The suppression is somewhat
more pronounced in the extreme case
$\sigma_{zz} \ll \sigma_{yy} \ll \sigma_{xx}$.

The reduction of the decoherence and heating rates is illustrated in
Fig. \ref{Fig:spinDecoherence} for the four cases discussed above.
We plot the spin decoherence rate~(\ref{Eq:spinDecoherenceRate}) for a
superposition state of the hyperfine levels
$\left| F=2,m_F=2 \right \rangle$ and $\left| F=2,m_F=1 \right \rangle$
in the ground state of $^{87}$Rb.
The anisotropic conductor is chosen such that the largest conductivity value
$\sigma_{xx}$ coincides with the one for Au, an isotropic conductor
taken as reference. The calculated rates for some specific materials are also given.
It can be seen that quasi-1D materials
are much more appealing to suppress heating and decoherence, although
about one order of magnitude can already be gained with layered
materials, even at a relatively small anisotropy.
In addition, for most anisotropic materials, even the best direction
conducts worse than Au; therefore the rates presented in this graph even for layered materials are
 smaller than for Au.

To conclude, heating and decoherence may be suppressed by several orders of magnitude even at room temperature, by using electrically anisotropic materials for current carrying structures on atom chips.

\section{Time independent spatial corrugations of the magnetic potential}
\label{Sect:fragmentation}
Anisotropic materials affect not only the magnetic noise,
leading to the time-dependent perturbations discussed so far, but also
the static corrugations of the trapping potential that fragment an
atom cloud. In this section, we discuss the current flow in
imperfect wires, described by a spatially modulated conductivity and
generalize the formalism developed in Refs. \cite{aigner,japha} to the
anisotropic case.

\subsection{System definition}
\label{Sect:sys_def}
\begin{figure}[t]
\vspace{1 cm}
\centering
\includegraphics[width=7.5cm]{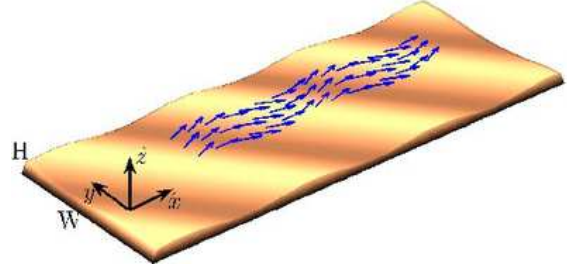}
\caption{Geometry of the current carrying wire. Planar conductivity fluctuations can arise from bulk conductivity changes or thickness variations. A single Fourier component $(k_x, k_y)$ is illustrated (color scheme represents conductivity changes) with the resultant variation of the current flow direction. The current is tilted toward low resistivity regions and perpendicular to the high resistivity ones. The conductivity fluctuation is characterized by an angle $\theta$, with $\tan(\theta)=k_x/k_y$. Measurement and calculation for the isotropic case found that the largest amplitude of transverse currents is formed along wave fronts oriented at $\theta = ±45^{\circ}$ . An intuitive explanation of this angle has been given in \cite{aigner}, and a formal model is developed in \cite{japha}.
  \label{Fig:geometry}}
\end{figure}

We now consider a thin metal wire with conductivity fluctuations, as
sketched in Fig. \ref{Fig:geometry}.  The coordinate system is again
defined such that the wire length $L$ is along the $\hat{x}$
direction, its width $W$ along $\hat{y}$, and its thickness $H$ along
$\hat{z}$. The anisotropic crystal used for the atom chip is again assumed
to be 'aligned' with the wire axis, as in Eq. (\ref{Eq:rhoDef}). This
is of course not the most general scenario, however it is sufficient
for our purposes here.

\subsection{Current corrugations}
 In the isotropic case \cite{aigner,japha} we start from Ohm's law
\begin{equation}
    \vec{E} = \rho \vec{J}
    \label{Eq:ohm_law},
\end{equation}
where $\rho = \sigma^{-1} = \rho_{0}  + \delta \rho$ is the scalar
resistivity comprised
of a homogeneous component $\rho_{0}$ and a fluctuation $\delta \rho$.
This is plugged into Maxwell's equation,
\begin{equation}
\nabla \times \vec{E} = 0 \label{Eq:maxwell}.
\end{equation}
We use a spectral expansion of
the fluctuation $\delta \rho (\vec{x})$ into plane waves
\begin{equation}
\delta\rho(\vec{x}) = \sum_{\vec{k}}{\rm e}^{{\rm i} \vec{k} \cdot
{\bf x} } \delta\rho( \vec{k} )
,
\end{equation}
where $\vec{k} = (k_x, k_y, k_z)$ is a
three-dimensional wave vector in which $(k_y, k_z)$ take the discrete
values $2\pi(m/W, n/H)$ with integers $m$ and $n$. Combining with the continuity equation $\nabla\cdot\vec{J} = 0$ we
obtain to first order in $\delta\rho$
\begin{equation}
\vec{\delta J}(\vec{k}) = \left(\frac{\vec{k}\cdot
k_x}{k^2}-\hat{x}\right)\frac{\delta\rho( \vec{k} )}{\rho_0}J_0,
\label{Eq:kIsoResult}
\end{equation}
per each $\vec{k}$ component.  Here $J_0 {\hat x}$ is the unperturbed current
density applied along the wire.
In the following, we focus on thin (flat) wires and neglect the $k_z$ component \cite{japha}.
For a fluctuation $\delta\rho( \vec{k} )$ with a given in-plane
wave vector $\vec{k} = k\left(\cos\theta, \sin\theta\right)$,
Eq. (\ref{Eq:kIsoResult}) implies a current fluctuation transverse to the
applied current that can be characterized by the ratio
\begin{equation}
\alpha^{\rm iso}( {\bf k} ) \equiv \frac{\delta J_y( {\bf k} )}{J_0} =
\frac{1}{2}\sin(2\theta)\frac{\delta\rho( {\bf k} )}{\rho_0}.
\label{Eq:isoResult}
\end{equation}
For small perturbations, $\alpha$ describes the angular amplitude of the
current deviation with respect to the $\hat x$-axis.  The main feature
is that this deviation is maximal when the wave vector $\vec{k}$ is
oriented at $\theta = \pm 45^{\circ}$, as illustrated in
Fig. \ref{Fig:geometry}.

The current density pattern is mapped onto the magnetic field measured by the atoms at a distance $d$ as
\begin{equation}
\beta^{\rm iso}({\bf k},d) = e^{-kd}\frac{1}{2}\sin(2\theta)\frac{\delta\rho( {\bf k} )}{\rho_0},
\label{Eq:beta}
\end{equation}
exhibiting the same angular dependence. This is observed in experiments by imaging the atoms' density profile. The
local column density profile is a direct measurement of the variations in the bottom of the trapping potential. The exponential term in $\beta$ acts as a filter such that short wavelengths compared to the atom-surface distance are suppressed in the magnetic field fluctuations spectrum.

In the case of electrically anisotropic materials, the resistivity
becomes a tensor $\rho_{kl} = \rho_{0,kl} + \delta \rho_{kl}$ that is inverse to Eq. (\ref{Eq:rhoDef}).
Maxwell's equation combined with Ohm's law then takes the
form
\begin{equation}
\left[\nabla\times\left(\hat{\rho}\vec{J}\right)\right]_i = \epsilon_{ijk}\partial_j\left(\rho_{kl}J_l\right) = 0,
\end{equation}
where $\epsilon_{ijk}$ is the Levi-Civita tensor and repeated indices
are summed over.

Rearranging and writing ${\bf J} = J_{0} {\hat x}$ for the unperturbed applied
current density, we obtain
\begin{equation}
\epsilon_{ijk}\rho_{0,kl}\partial_j J_l =
\epsilon_{ikj}\left(\partial_j\delta\rho_{kx}\right)J_0.
\end{equation}
Again focusing on the 'aligned' case as in Eq. (\ref{Eq:rhoDef}), and combining with the continuity equation $\partial_i J_i = 0$, we solve
the set of equations to get the components of the current
fluctuations,
\begin{equation}
\delta \vec{J}^{\rm aniso}( \vec{k} )  =
    \left(\frac{k_x\, \vec{q} }{ \vec{k}\cdot\vec{q} } - \hat{x}\right)
\frac{ \delta \rho_{x}( \vec{k} ) }{ \rho_{0,x} }J_0,
\label{Eq:k_result_aligned}
\end{equation}
where $\vec{q} = \left(k_x/\rho_{0,x}, k_y/\rho_{0,y}, k_z/\rho_{0,z}\right)$
and $\delta \rho_{x}( \vec{k} ) / \rho_{0,x}$ is the
relative fluctuation of the resistivity along the $\hat{x}$-direction.
This reduces to the isotropic result (\ref{Eq:kIsoResult}) when
$\rho_{0,x} = \rho_{0,y} = \rho_{0,z}\equiv \rho_0$.

Focusing again on wave vectors $\vec{k} =
k\left(\cos \theta, \sin \theta\right)$ and a thin
wire, we get
\begin{equation}
\alpha^{{\rm aniso}} = \frac{
{\textstyle\frac12}\sin 2\theta }{\sin^2\theta + r\,\cos^2\theta }
\frac{\delta \rho_x}{\rho_{0,x}},
\label{Eq:ang_result_aligned}
\end{equation}
where the anisotropy ratio is $r \equiv \rho_{0,y}/\rho_{0,x}$. Here it is evident that the orientation of the current wave fronts is different from the isotropic case.  While in
the isotropic case, $r = 1$, we recover the $\sin(2\theta)$ dependence
of Eq. (\ref{Eq:isoResult}), the angular dependence will asymptote to
$\tan(\theta)/r$  in the highly anisotropic limit, $r \gg 1$.
			
\subsection{Analysis}
\label{Sect:analysis_aligned}
\begin{figure}
\centering
\vspace{0.5cm}
\includegraphics[width=\columnwidth]{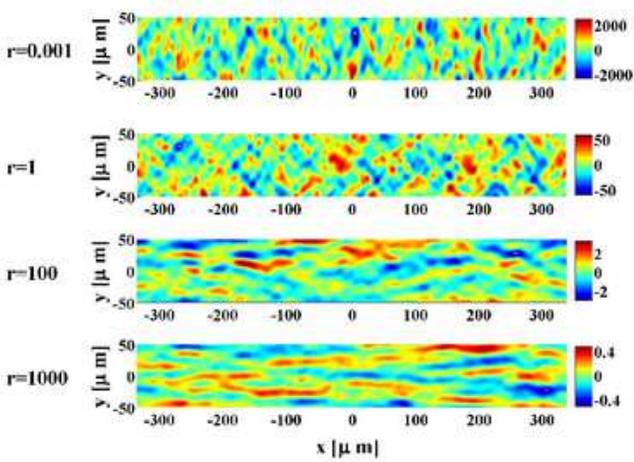}
\caption{Comparison of magnetic corrugation patterns for different anisotropy ratios $r = \sigma_{xx} / \sigma_{yy}$. The electron flow is mapped onto the magnetic field, shown here in false color (arb. units). Red colors indicate large magnetic field rotation $\beta^{{\rm aniso}}\left(x,y,z = d\right)=e^{-kd}\alpha^{\rm aniso}( \vec{k} )$ (Eq. (\ref{Eq:ang_result_aligned})), and hence large transverse currents. As the anisotropy is larger, the value of the preferred angle $\theta_{{\rm max}}$ is shifted, following the law $\tan \theta_{\rm max} = r^{1/2}$, and the overall signal intensity is suppressed, scaling on average as $r^{-3/4}$. For the opposite case, $r \ll 1$ (badly conducting axis along the wire), the situation is reversed, and the fragmentation is enhanced.
 \label{Fig:widthscan_aligned}}
\end{figure}

Let us introduce the angle $\theta_{\rm max}$ as that corresponding to
the wave vector $\vec{k}$ producing the largest current deviation.
A simple calculation shows that this preferred angle satisfies $\tan \theta_{\rm max} = r^{1/2}$. In the highly anisotropic case ($r \gg 1$), the current corrugation thus has wave fronts essentially perpendicular to the applied
current (Fig. \ref{Fig:widthscan_aligned}, lower panels). In addition, the peak corrugation amplitude, taken at the preferred angle
and normalized to the isotropic case, scales like
$\alpha^{\rm aniso}(k,\theta_{{\rm max}}) / \alpha^{\rm iso}(k,45^{\circ})
\propto r^{-1/2}$.
However, in an experiment, one rather observes the angle-averaged
power spectrum for which we find a scaling
\begin{equation}
\left(
\frac{
    \int{{\rm d}\theta \left|
\alpha^{\rm aniso}\left(k,\theta\right)\right|^2}
}{
\int{ {\rm d}\theta
\left|\alpha^{\rm iso}\left(k,\theta\right)\right|^2}
}
\right)^{1/2} \propto r^{-3/4}
\label{Eq:rms-spectrum-scaling}
\end{equation}
in the high anisotropy limit $r \gg 1$.
Highly anisotropic materials for trapping wires hence have the advantages of
controlling the current scattering pattern in the wire and of
suppressing fragmentation quite strongly.

In Fig. \ref{Fig:widthscan_aligned} we show simulations of the
fluctuations of the magnetic potential $\beta^{{\rm aniso}}\left(x,y,z = d\right)=e^{-kd}\alpha^{\rm aniso}( \vec{k} )$,
in arbitrary
units. The plots are made in false color and consider different anisotropy ratios. We assumed a white spectrum for $\delta \rho( \vec{k} )$.
The simulated wires are of width $W=200 \mu m$, thickness $H
= 2 \mu m$, and the observation point (atom-surface distance) is at $d =
3.5 \mu m$.  The scan is across the central 100x680$\mu m^2$, safely
away from the wire edge.  The change in orientation is clearly
seen when the anisotropy ratio is varied. Note also the numerical
values of the field corrugation amplitude (color bar), from which the overall suppression of fragmentation can be inferred.

\section{Materials and experimental design considerations}
\label{Sect:mat_fab}
In this section we discuss issues pertaining to the experimental verification of the theory, as well as possible applications to quantum technology devices. We start with a review of electrically anisotropic materials which may be considered as potential candidates for atom chips, and then consider the effect of applying current to the wire, and hence heating it, on thermal noise properties. We conclude by discussing the modifications imposed by the need to add longitudinal confinement to the trapping guide, in the form of a z-shaped wire.

\subsection{Electrically anisotropic materials}
\label{Sect:mat}

Electrically anisotropic materials have been studied mostly in the context of characterizing their electrical transport properties (e.g. as a function of temperature or fabrication methods and parameters), or their magnetic properties. A large portion of these materials are also high-$T_c$ superconductors, hence they are also interesting in that context. Table 1 shows the relevant properties of some anisotropic materials.

\begin{table*}
	\centering
		\begin{tabular}{|c|c|c|c|c|c|c|c|c|}
		\hline
 			 Material Class & Material & $\frac{\rho_a}{\rho_{Au}}$ & $\frac{\rho_c}{\rho_a}$ &  $\frac{\rho_c}{\rho_b}$ &  $\frac{\rho_a}{\rho_b}$ & Magnetic properties  & Source \\
  		\hline
  		\multirow{3}{*}{hcp metals} & ${\rm Sc}$ & 27.6 & 0.37 & & & Paramagnetic & \cite{TableSc} \\
   		& ${\rm Ga}$ & 7.8 & 3.21 & 7.07 & 2.2 & Diamagnetic& \cite{TableGaTe} \\
  		& ${\rm Te}$  & 7000 & 3.64 & & & Diamagnetic&  \cite{TableGaTe} \\
  		\hline
  		\multirow{4}{*}{Layered compounds} & ${\rm LaSb_2}$ & 65 & 16.55 & & & Paramagnetic & \cite{TableLaSb2} \\
  		  & ${\rm Sr_2RuO_4}$ & 180-450 & 50 - 300 & & & Paramagnetic  & \cite{TabelSr2a,TabelSr2b,TabelSr2c,TableSr2d}\\
  		 & ${\rm Sr_3Ru_2O_7}$ & 3.8 & 23.5 & & & & \cite{TableSr3} \\
  		  & ${\rm NaCo_2O_4}$ & 86 & 42.11 & & & & \cite{TableNaCoO}\\
  		 \hline
  		 \multirow{2}{*}{Ladder-spin compounds} & ${\rm Ca_{14}Cu_{24}O_{41}}$ & 9$\cdot10^5$ & 0.01 & & & & \cite{TableCaCuO} \\
  		  & ${\rm Sr_3Ca_{11}Cu_{24}O_{41}}$ & 4500& 0.1 & $10^{-5}$ & $10^{-4}$ & Paramagnetic  & \cite{TableSr3Ca} \\
  		 \hline
  		 \multirow{3}{*}{Cuprates} & ${\rm Bi2212}$ & 180 & $1.5\cdot10^4$ & & & Paramagnetic& \cite{TableBi2212}\\
  		  & ${\rm YBCO}$ & 90-360 & 25-60 & & & Paramgnetic& \cite{TableYBCO} \\
  		  & ${\rm La_{2-x}Sr_xCuO_4}$ & 45-400 & 100 - 360 & & & Paramgnetic & \cite{TableLaSrCuO} \\
  		 \hline
  		 \multirow{2}{*}{Graphites} & Natural single crystal & 25.8 & 105 & & & Nonmagnetic  & \cite{graphite_single_crystal} \\
  		  & ${\rm HOPG}$ & 18 & 3750 & & & Nonmagnetic  & \cite{TableHOPG,TableHOPG2,TableHOPG3} \\
  		 \hline
  		 \multirow{2}{*}{Perovskites} & ${\rm LaTiO_{3.41}}$ & 362 & $8.75\cdot10^4$ & 850 & $10^{-2}$ & Paramagnetic& \cite{TableLaTiO} \\
  		   & ${\rm SrNbO_{3.41}}$ & 226 & $3\cdot10^3$ & 50 & $2\cdot10^{-2}$ & Diamagnetic& \cite{TableSrNbO} \\
  		 \hline
  	\end{tabular}
  	  \caption{Electrically anisotropic materials as potential candidates for atom chip wires. The indices $a$, $b$, and $c$ denote the crystal axes. We assume the good conductivity to be along the wire, and the worst to be perpendicular to the wire axis, in the plane parallel to the substrate (see definitions in Sect. \ref{Sect:rates}).}
  \label{Tab:tablecaptionis}
\end{table*}

Although some hexagonal metals exhibit electrical\linebreak anisotropy, it is usually smaller than $r=4$. Higher\linebreak anisotropy ratios can be found in several layered compounds, as well as in some ladder-spin compounds, also shown in the table. A potential problem using these compounds is a strong dependence of electrical and magnetic properties on the exact relative composition, which could result in spatial inhomogeneity of the material properties within the wire. Paramagnetic materials are also considered less preferable for magnetic traps. It should be noted however, that atom traps with permanent magnets have also been studied \cite{permanent_magnets}. Some materials may have problems being in an ultra-high vacuum (UHV) environment, either due to out-gassing, or to a degradation of their electrical properties as the oxygen content in the ambient atmosphere is reduced (e.g. YBCO \cite{YBCO_UHV}).

We note here especially different types of graphite, which seem to be attractive candidates for atom chips. Natural single crystal graphite \cite{graphite_single_crystal}, exhibits an anisotropy of $r \approx 100$ and is non-magnetic. Highly ordered pyrographite (HOPG) is a form of synthetic single crystal graphite, which exhibits a very high (layered) anisotropy of  $r > 3500$. This comes without significantly hindering of the good conductance along the wire, which is comparable for example to that of Ti, only one order of magnitude lower than Au or Cu. However, the combination of graphite being a layered material with its softness creates a challenge in fabricating wire-sized structures out of it. Cutting the material perpendicular to the layers' plane may result in breaking the material into its layers (typical layer thickness on the order of $\sim\! 30$ nm for HOPG \cite{TableHOPG3}). This has importance given that the anisotropy is perpendicular to the layers' plane.

In Table 1 we also list as examples two materials with quasi-1D conductance. The Perovskite-type transition-metal oxides ${\rm SrNbO_{3.41}}, {\rm LaTiO_{3.41}}$ present relatively high conductivity in one direction, only one or two orders of magnitude below regular metals and much better than typical semiconductors. These are highly quasi-1D materials, as in fact $\sigma_{zz} \ll \sigma_{yy} \ll \sigma_{xx}$. These materials are commonly fabricated by the zone-melting method which enables an extremely high precision in the material composition. This precision is required as materials with close structural proximity but different oxygen content have quite different electrical properties. In this method, high purity compound powders are combined at a high temperature through a series of chemical reactions \cite{Lichtenberg1}. Special care is required to maintain exact weights of the various components and the desired oxygen content. The composite compounds are then pressed and sintered into two rods. These are subjected together to a melting process, limited to a small region, in which one rod acts as a feed and the other as a seed. The melting zone is slowly translated, and following the solidification of the melted parts, high-quality crystals can be obtained. For extensive reviews on these materials, including fabrication aspects, see \cite{Lichtenberg1,Lichtenberg2}. The common size of the crystals formed by this method is large compared with the requirements for atom chip micro-structures. However the electrical properties are maintained even upon size reduction by means of applying pressure \cite{Lichtenberg3,Lichtenberg4} and various cutting methods. Still, in order to obtain the requirements for micro-structures more delicate methods, such as ion-beam etching, may be needed. These materials could be contacted electrically by standard means. To conclude, there seem to be no inherent limitations in obtaining structures adequate for atom chips.

\subsection{Influence of wire heating on noise measurements}
\label{Sect:fab_noise}

\begin{figure}
\centering
\includegraphics[width=\columnwidth]{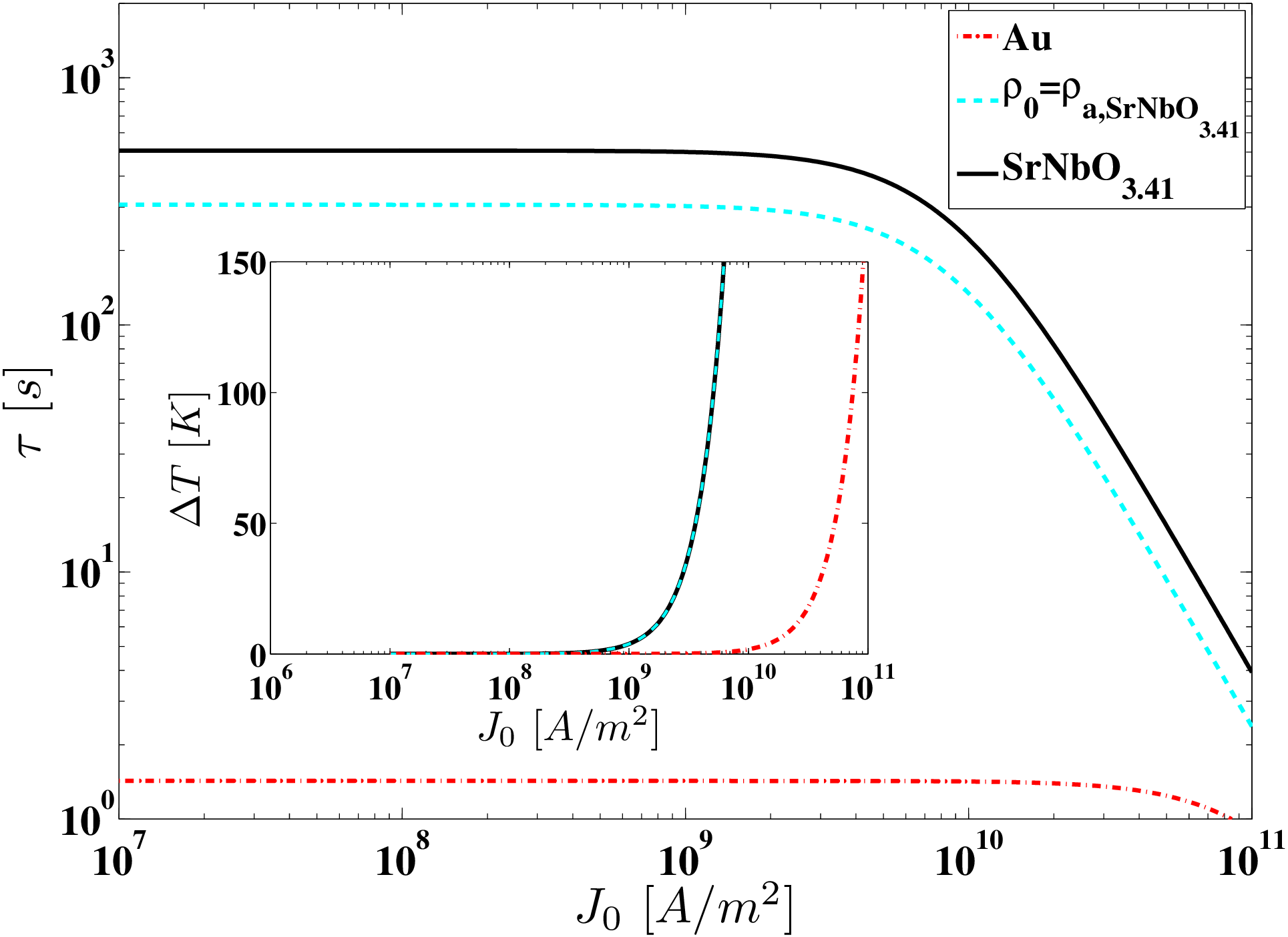}
\caption{Drawback of electrically anisotropic materials: reduced magnetic trap lifetime as a current-carrying wire heats up. The trap lifetime due to thermal noise induced spin flips is plotted as a function of current density $J_{0}$. For $J_{0} \to 0$, the level of pure thermal noise (at $T = 300\,{\rm K}$) is reached. The lifetime of metal wires (Au, solid red) is not significantly affected up to $J_0 \approx 10^{11} A/m^2$. For high resistivity materials relative to Au the lifetime drops significantly as the current density is increased above $10^9 A/m^2$, due to wire heating. The lifetime of anisotropic materials (SrNbO$_{3.41}$, solid black) is improved by a factor $\sim\! 2$, and their current density dependence is the same as that of isotropic materials of equal conductivity along the wire (dashed blue), (Sect. \ref{Sect:TrapLoss}). Inset: Temperature rise (above 300 K) of the wire as the external current density is increased. Wire parameters are $W=10 \mu m$, $H=2.15 \mu m$, and trap distance $d = 5 \mu m$. This calculation follows the model developed by Groth \itshape et al.\upshape ~\cite{HDheating}. The substrate was assumed to be made of Si, with a 20 nm thick insulation layer of ${\rm SiO_2}$. Steady state of the temperature increase was assumed to be reached after $t=30 s$. The resistivity was assumed to be independent of temperature in the $\Delta T$ range shown here, up to 150$^{\circ}$ above room-temperature \cite{ValeryAlloys,TableSrNbO}. \label{Fig:tauCurrentDeltaT}}
\end{figure}

An important practical issue that may compromise the reduction of thermal noise is the heating of wires when current is applied. In principle, to investigate thermal noise itself, no current is needed in the probed structure; however, considering an actual application with a magnetic micro-trap, there will be current in the wires, and hence heating of the surface. As can be seen from the prefactor of the magnetic field power spectrum (\ref{Eq:powerSpectrumFull}), this heating would increase the noise, approximately linearly in the surface temperature. Bearing in mind that most anisotropic materials have a lower conductivity than metals, this implies that running a current through an anisotropic material wire would hinder the improved lifetimes and coherence times. In Fig. \ref{Fig:tauCurrentDeltaT} we plot as an example the expected lifetime as a function of current density $J_0$ along the wire, taking into account wire heating according to the model developed by Groth \itshape et al. \upshape \cite{HDheating}. As the relevant component to the heating process is the applied current, there is no difference between an anisotropic material and an isotropic one with the same conductivity along the wire, up to a factor $\sim\! 2$ due to the anisotropy (Sect. \ref{Sect:TrapLoss}). The longer lifetime relative to the Au case is mostly due to the lower conductivity relative to Au (in the direction along the wire for anisotropic materials), up to current densities of $10^9 {\rm A/m}^2$. For higher current densities we see that materials with low conductivity heat up much more than metallic ones (shown in the inset), and the corresponding lifetime drops by two orders of magnitude for current densities of up to $10^{11} {\rm A/m}^2$, which is, however, already a fairly high value for small metallic wires \cite{HDheating}. It should be noted that the prefactor including the surface temperature in Eq. (\ref{Eq:powerSpectrumFull}) in the context of trap loss is the same for atoms' heating- and decoherence-rates, hence this plot is relevant also for these processes.

The conclusion here is that although electrically\linebreak anisotropic materials are promising candidates for reducing the coupling of noise to the atoms, their advantages are limited to moderate current densities. This should not pose overly stringent restrictions on possible applications, as at small atom-surface distances the required currents for small and tight traps are not very high.

\subsection{Contribution from structures providing longitudinal confinement}
\label{Sect:desgin3D}
\begin{figure}
\centering
\includegraphics[width=\columnwidth]{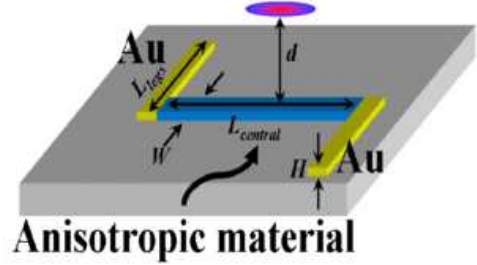}
\caption{Schematic illustration of the studied z-shaped wires. The atoms are trapped at a distance $d$ from the central part of the wire of length $L_{central}$, made of an anisotropic material as summarized in Table 2. This part of the wire provides the radial trapping potential. The longitudinal confinement is provided by the z-wire 'legs' of length $L_{legs}$, made of Au. The width $W$ and thickness $H$ are the same for the whole wire.
\label{Fig:3DIllustration} }
\end{figure}

The most common Ioffe-Pritchard magnetic micro-trap configuration is that of a z-shaped wire combined with an external bias field \cite{ronReview}. The central part of the wire creates a guide potential, while the 'legs' provide the longitudinal confinement as well as the non-zero field value at the trap minimum, as illustrated in Fig. \ref{Fig:3DIllustration}. In realistic magnetic micro-traps, where longitudinal-confinement is required, considering only the potential generated by the central part of the z-wire does not suffice for the anisotropic case, as the contribution from the 'legs' of the z-wire has to be taken into account. This is expected to be more and more important as the miniaturization of the micro-structures advances toward high packing density needed for example for matter-wave quantum technology devices. To asses the influence of the 'legs', we first consider a relatively large wire (atom-surface distance $d=50\mu m$, wire width $W=100\mu m$, thickness $H=1\mu m$, central part length $L_{central} = 1 mm$, and 'legs' length $L_{legs}=1.5mm$), representing a typical micro-trap. To demonstrate the scenario of highly dense traps, we also consider the geometry of a much smaller structure ($d=3\mu m$, $W=6\mu m$, $H=1\mu m$, $L_{central} = 20\mu m$, and $L_{legs}=30\mu m$). For both geometries we compare the spin decoherence rate as a function of material of the central part of the wire, normalized to that of a Au structure. We also take into account a current density $J_0$ applied to the wire, limiting the heating of the wire to $\Delta T \approx 100^{\circ}C$ (Sect. \ref{Sect:fab_noise}).

The results are summarized in Table 2. Obviously it would be preferable to have the entire wire made of the anisotropic materials. However, for that the orientation of the crystalline axis in the area forming the 'legs' should be perpendicular to the central part of the z-wire, increasing fabrication complexity. However, having the 'legs' made of a normal metal, with isotropic conductivity, simplifies the fabrication, lowers power consumption and wire heating. Indeed we find that even if the metal 'legs' to the anisotropic material reach up to the central part of the trapping wire, the reduction of the the decoherence rate is still significant.

\begin{table}
\centering
		\begin{tabular}{|c|c|c|c|}
  		\hline
  		\multicolumn{2}{|c|}{Wire Material} & Large Wire & Small Wire \\ 
  		\hline
  		Central part  & Legs & $\gamma_{{\rm spin~ decoherence}}$ & $\gamma_{{\rm spin~ decoherence}}$ \\
  		 \hline
  		 ${\rm Au}$ & ${\rm Au}$ & 1 & 1 \\
  		 \hline
  		 ${\rm SrNbO_I}$ & ${\rm Au}$ & 3.68$\cdot 10^{-4}$ & 9.9$\cdot 10^{-3}$ \\
  		 \hline
  		 ${\rm SrNbO_{II}}$ & ${\rm Au}$ & 4.39$\cdot 10^{-4}$ & 1$\cdot 10^{-2}$ \\
  		 \hline
  		 ${\rm HOPG}$ & ${\rm Au}$ & 9.7$\cdot 10^{-3}$ & 1.8$\cdot 10^{-2}$ \\
  		 \hline
  	\end{tabular}
  	  \caption{The affect of longitudinal confinement on the reduction of the decoherence-rate when using electrically anisotropic materials. Two z-shaped wire geometries, as illustrated in Fig. \ref{Fig:3DIllustration}, are considered: a large wire ($d=50\mu m$, $W=100\mu m$, $H=1\mu m$, $L_{central} = 1 mm$, and $L_{legs}=1.5mm$), and a small wire ($d=3\mu m$, $W=6\mu m$, $H=1\mu m$, $L_{central} = 20\mu m$, $L_{legs}=30\mu m$). The good conductivity axis (the $a$-axis) is along the central part of the wire. The spin decoherence rate, normalized to that of a structure entirely made of Au, is calculated for different combinations of wire material compositions. The externally applied current density was set to $3.3\cdot10^9 {\rm A/m}^2$ and $8.5\cdot10^9 {\rm A/m}^2$ for the large and small wires, respectively, such that the heating of the wire would not exceed 100$^\circ$C (see section \ref{Sect:fab_noise}). The difference between the two calculations for SrNbO$_{{\rm 3.41}}$ is whether the worst conductivity axis of this quasi-1D material is along the wire width (SrNbO$_{{\rm I}}$) or its thickness (SrNbO$_{{\rm II}}$). Even for small structures, with the Au 'legs' very close to the trapping area over the central part of the z-wire, the suppression of the decoherence and heating rates is still significant, further showing the appeal of using electrically anisotropic materials for realistic tight traps.}
  \label{Tab:2}
\end{table}

\section{Conclusions}
We have shown that utilizing electrically anisotropic materials on an atom chip will significantly suppress harmful mechanisms due to atom-surface interaction. Heating- and decoherence-rates due to thermal noise from the surface are expected to be strongly reduced even at room-temperature, scaling linearly with the electrical anisotropy for certain types of anisotropic materials. This is maintained also in the presence of an external current applied to the wire (causing wire heating), up to current densities on the order of $10^{9} A/m^2$. We have shown that structures perpendicular to the wire axis, needed for longitudinal confinement, do not significantly hinder the improvements in heating and \linebreak decoherence-rates, even when made of normal metal. The expected trap lifetime is not substantially modified by the use of such materials, although a significant improvement can be achieved by cooling the surface to low temperatures.
Potential corrugations due to time independent current scattering patterns in the imperfect current carrying wires is expected to be suppressed with the anisotropy $r$ as $r^{-3/4}$ on average. The preferred orientation of scattered current wave fronts can also be controlled over a wide range by varying the electrical anisotropy, becoming perpendicular to the applied current in the extreme anisotropic case. Specific materials have been suggested as potential candidates for implementing the results obtained in this work. This could lead to improved functionality of atom chip based applications such as clocks, sensors and quantum information communication and processing.

\begin{acknowledgement} 

We thank the team of the Ben-Gurion University Weiss Family Laboratory for Nanoscale Systems (www.bgu.ac.il/nanofabrication). T. D. thanks F. Lichtenberg for details on perovskite-related materials and their fabrication. We gratefully acknowledge the support of the European Union 'atomchip' (RTN) collaboration, the German Federal Ministry of Education and Research (BMBF-DIP project), the American-Israeli Foundation (BSF) and the Israeli Science Foundation.
\end{acknowledgement}


\begin{thebibliography}{999}
\bibitem{DecoherenceNote}
Indeed, \cite{Reichel_clock} is the only experiment in which surface induced spin decoherence was measured. However, as the main objective of this experiment was to minimize shifts to an on-chip atomic clock, a specific choice of superposition states and magnetic field value was made, resulting in a surface noise induced decoherence a level weaker than that induced by the technical instabilities of the experiment. In regards to spatial decoherence, no experiment to study the dependence of the decoherence rate on atom-surface distance was performed as of yet.
\bibitem{Note:TrapInhomogeneity}
It should be noted that considering the noise only at the trap center (as usually done) may not be sufficient, as by doing so one neglects the fact that the trap is commonly spatially inhomogeneous (e.g. harmonic). The atoms are distributed in the trap with a certain density profile, and move as they have finite temperature. Taking this into account introduces corrections to the theory which in some cases may be important. However, for the purposes of this study this is not crucial, and we will address this issue separately.
\end{thebibliography}
\end{document}